\documentclass[conference, a4paper]{IEEEtran}
 \IEEEoverridecommandlockouts
\ifCLASSINFOpdf
\else
   \usepackage[dvips]{graphicx}
\fi
\usepackage{url}

\usepackage{cite}
\usepackage{graphicx,amsmath,amssymb,tikz,psfrag}
\usepackage{color}
\usepackage{fancyhdr} 
\usepackage{tikz}
\usetikzlibrary{shapes,arrows,chains}
\usepackage{xcolor}
\usepackage{algorithm}
\usepackage{algpseudocode}
\usepackage{enumerate}
\usepackage{diagbox}

\begin{document}

\title{Secure Rate-Splitting for MIMO Broadcast Channel with Imperfect CSIT and a Jammer 
 
}
\author{\IEEEauthorblockN{Tong Zhang\textsuperscript{$1$}, Dongsheng Chen\textsuperscript{$1$},  Na Li\textsuperscript{$2$}, Yufan Zhuang\textsuperscript{$1$}, Bojie Lv\textsuperscript{$1$},     and Rui Wang\textsuperscript{$1$}}

	\IEEEauthorblockA{\textsuperscript{$1$}Department of Electrical and Electronic Engineering, Southern University of Science and Technology, China }
 	\IEEEauthorblockA{\textsuperscript{$2$}National Engineering Lab for Mobile Network Technologies, Beijing University of Posts and Telecommunications, China} 
	Email:
\{zhangt7, 11712127,  zhuangyf2019, lyubj, wang.r\}@sustech.edu.cn, 
Lina-Lena@bupt.edu.cn 

}

\maketitle

\begin{abstract}
In this paper, we investigate the secure rate-splitting for the two-user multiple-input multiple-output (MIMO) broadcast channel with imperfect channel state information at the transmitter (CSIT) and a multiple-antenna jammer, where each receiver has an equal number of antennas and the jammer has perfect channel state information (CSI). Specifically, we design a  secure rate-splitting multiple-access strategy, where the security of split private and common messages is ensured by precoder design with joint nulling and aligning the leakage information, regarding different antenna configurations. Moreover, we show that the sum-secure degrees-of-freedom (SDoF) achieved by secure rate-splitting is optimal and outperforms that by conventional zero-forcing. Therefore, we reveal the sum-SDoF of the two-user MIMO broadcast channel with imperfect CSIT and a jammer, and validate the superiority of rate-splitting for the security purpose in this scenario with emphasis of MIMO.
\end{abstract}

\begin{IEEEkeywords}
 Broadcast channel, MIMO, rate-splitting, imperfect CSIT, secure degrees of freedom
\end{IEEEkeywords}


\section{Introduction}

For the 6G cellular networks, the rate-splitting (RS) is a prospective solution for multi-user (massive) multiple-access  technology, due to its nature of non-orthogonal transmission and robust interference management \cite{1,2,3}.
It was shown in \cite{203,260} that RS can achieve the maximal multiplexing gain in multiple-input multiple-output (MIMO) networks with imperfect channel state information at the transmitter (CSIT), where the maximal multiplexing gain indicates the maximal number of interference-free data streams and is recognized as the name  ``degrees-of-freedom'' in open literature. Likewise, the secure degrees-of-freedom (SDoF) stands for how many interference-free data streams with  security guarantee can MIMO networks afford \cite{100,101,103,102,104}. The secure RS (S-RS) is able to attain optimal SDoF, where a couple of initial results were disclosed in \cite{50}.  Although the design of \cite{8} is SDoF-optimal, \cite{8} is only limited to the $K$-user multiple-input single-output (MISO) broadcast channel with imperfect CSIT and a jammer. Motivated by the SDoF optimality in the $K$-user MISO broadcast channel, we investigate the S-RS in the two-user MIMO broadcast channel. 

Recently, RS has attracted a plenty of research interests. 
 It was shown in \cite{252} that power partitioned rate-splitting multiple-access (RSMA)
	achieves the optimal DoF in an overloaded
	MISO broadcast channel with heterogeneous CSIT qualities. It was proven in \cite{253} that a RS-based design achieves higher max-min
 DoF compared with conventional
	No RS designs. In \cite{254}, the optimal DoF Region of the $K$-User MISO broadcast channel with imperfect CSIT was achieved by RS. For $K$-cell MISO interference channel with an arbitrary CSIT quality of each interfering link, \cite{255} identified the DoF region achieved
	by RS. Moreover, the applications of RS were discussed in integrated sensing and communication (ISAC)  systems \cite{4}, intelligent reflecting surface (IRS) networks \cite{5}, satellite systems \cite{7}, and so on.

Due to the broadcast nature of wireless medium, the transmission signal may be wiretapped by eavesdroppers, thus arousing the security issue of wireless communications \cite{200,201,202}. The solutions to  security issue in RS has been studied in \cite{21,22,23,24,25,50}. To ensure the security of RS, the authors in \cite{21} first considered the dual use of common message transmission and jamming functionalities in the two-user MISO broadcast channel with an external eavesdropper. In \cite{22}, the  RS was designed in secure unmanned areal vehicle (UAV) networks. In \cite{23}, the secure sum-rate is maximized in the two-user MISO broadcast channel with an external eavesdropper. The adaptive beamforming strategy and power allocation was investigated in \cite{24} two-user MISO broadcast channel with an external eavesdropper. The authors in \cite{25} applied RS in polar codes for communication over a multiple-access wiretap channel with two transmitters under strong secrecy. In \cite{50}, for the $K$-user MISO broadcast channel with imperfect CSIT and a multiple-antenna jammer, a S-RS design was proposed to achieve an exceptional sum-SDoF. However, all of the above works have not studied the MIMO broadcast channel scenario.  Note that \cite{250,251} investigate  RSMA in MIMO broadcast channel with multiple antennas at each receiver. But security is not considered. Therefore, the role of number of receive antennas on the secure rate splitting is still a research problem.

In this paper, motivated by the research pitfalls, we investigate the S-RS for the  two-user MIMO broadcast channel with imperfect CSIT and a multiple-antenna jammer, where each receiver has equal number of antennas and the jammer has perfect CSIT.   To the best of our knowledge, this is the first work that investigates the SDoF of secure rate-splitting in MIMO broadcast channel with multiple receive antennas. Specifically, we design transmit precoders to enable the S-RS by joint nulling and aligning the leakage information, regarding to different antenna configurations. Since the spatial benefits of MIMO settings are exploited, we show that the sum-SDoF achieved by S-RS is optimal and outperforms that by conventional zero-forcing, in the  two-user broadcast channel with imperfect CSIT and a jammer. 

\textit{Notations}: Matrices and vectors are represented by upper and lowercase
boldface letters, respectively. The operator $\|\cdot\|$ and $\mathbb{E}\{\cdot\}$ stand for Frobenius  norm and expectation. $\log$ refers to $\log_2$. $\mathcal{O}(A)$ represents the same order of $A$.



\section{System Model}

We consider a two-user MIMO broadcast channel with confidential messages (BCCM), where a  $M$-antenna transmitter aims to deliver two independent and confidential messages, to $N$-antenna receivers 1 and 2, respectively, as depicted in Fig. 1. The message desired by receiver $i$ is denoted by $W_i$. Besides, there is a jammer to assist the transmitter. The jammer is equipped   with $J$ antennas, where $J  \ge 2N$. The channel between the transmitter and the receiver  $i$ is denoted by $\textbf{H}_i \in \mathbb{C}^{N\times M}$. The channel between the jammer and the receiver $i$ is denoted by $\textbf{G}_i\in \mathbb{C}^{N\times J}$. The transmit signal\footnote{We will show the proposed S-RS design of transmit signals later on.} of the transmitter is denoted by $\textbf{x}_t \in \mathbb{C}^M$. The transmit signal of the jammer is denoted by $\textbf{x}_h \in \mathbb{C}^M$. Thus, the received signal  at the receiver $i$ can be written as 
\begin{equation}
	\textbf{y}_i = \textbf{H}_i\textbf{x}_t + \textbf{G}_i\textbf{x}_h + \textbf{n}_i,
\end{equation}
where the additive white Gaussian noise (AWGN) at receiver $i$ is denoted by $\textbf{n}_i$, which follows complex Gaussian distribution  $\mathcal{CN}(\textbf{0},\textbf{N}_i)$ with positive definite matrix $\textbf{N}_i$. The total transmit power of the transmitter and jammer is bounded by a maximum power $P_{\max}$, which is written by $\mathbb{E}\{\|\textbf{x}_t\|_2^2\} + \mathbb{E}\{\|\textbf{x}_h\|_2^2\} \le P_{\max}$. Due to imperfect channel estimation, it is reasonable to assume that the CSIT is imperfect. The jammer is assume to be more powerful than the transmitter, thus it has perfect CSI. According to \cite{1,2,3}, we have the following imperfect CSIT model: 
\begin{equation}
	\textbf{H}_i = \hat{\textbf{H}}_i + \tilde{\textbf{H}}_i,
\end{equation}
where $\hat{\textbf{H}}_i$ denotes the estimate CSI, and $\tilde{\textbf{H}}_i$ denotes the unknown complement of estimate CSI. Since the accuracy of channel estimation scale with signal-to-noise ratio (SNR), $\|\tilde{\textbf{H}}_i\|^2$ scales as $\mathcal{O}(P^{-\alpha})$, where $\alpha \in [0,1]$. It can be seen that $\alpha = 1$ represents perfect CSIT case, while $\alpha = 0$ represents no CSIT case. We assume that the CSI at the receivers is perfect.

\begin{figure}[t]
	\centering
	\includegraphics[width=2.3in]{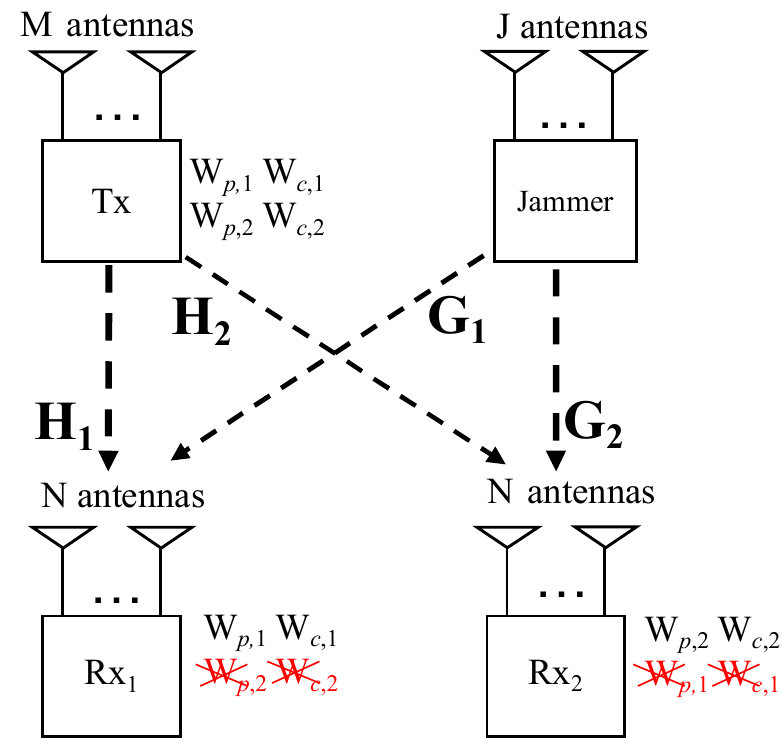}
	\caption{Illustration of two-user MIMO broadcast channel with a jammer, where the  confidential message $W_i,\,i=1,2$ is splitted into $W_{p,i},W_{c,i}$.}
\end{figure}

Denote the SNR as $\rho$. A code $\{2^{nR_i(\rho)},i=1,2,3\}$ with  achievable rate $R_i(\rho)$ and $n$ channel uses is defined below. A stochastic encoder at the transmitter, denoted by $f(\cdot)$, encodes the confidential messages to an transmit signal based on $\{\hat{\textbf{H}}_i\}$. A jammer transmits artificial noise signal based on perfect CSI. A stochastic decoder at the receiver $i$, denoted by $g_i(\cdot)$, decodes the estimated desired confidential messages using CSI at receivers and received signals. Let us denote the estimated  messages by $\hat{W}_1,\hat{W}_2$. The reliability constraint is given by
\begin{equation}
	\text{Pr}[\hat{W}_i \ne W_i] \le \epsilon_n,
\end{equation}
where $\epsilon_n$ approaches $0$ as $n$ goes to infinity. Also, the code needs to satisfy the
weak security constraint, given by
\begin{subequations}
	\begin{eqnarray}
		I(W_1,\mathcal{Y}_2^n)/n \le \epsilon_n, \\
		I(W_2,\mathcal{Y}_1^n)/n \le \epsilon_n,
	\end{eqnarray}
\end{subequations}
where the assemble of received signals of receiver $i$ across $n$ channel use is denoted by $\mathcal{Y}_i^n,i=1,2$. The sum of secure capacity is defined as $C_s = \max (R_1(\rho) + R_2(\rho))$. The sum-SDoF, denoted by $d_1 + d_2$, is the
first-order approximation of sum secure capacity in high SNR
regime and defined as
\begin{equation}
	d_1 + d_2 \triangleq \lim_{\rho \rightarrow {\cal{1}}} \frac{C_s}{\log \rho}.
\end{equation}

\section{Main Results and Discussion}

 \textbf{Theorem 1}: For the two-user MIMO BCCM with imperfect CSIT and a jammer, defined in Section-II, the sum-SDoF lower bound achieved by the S-RS is given as follows:
 \begin{equation}
 d_1 + d_2 \ge  \begin{cases}
 M, & \frac{M}{N} \le 1, \\
 N + \alpha(M-N), & 1 < \frac{M}{N} \le 2, \\
 N(1+\alpha), & 2 < \frac{M}{N}.
 \end{cases} \label{Result1}
 \end{equation}
 
 \begin{IEEEproof}
 	Please refer to Section-IV.
 \end{IEEEproof}
 
 To examine the power of S-RS, we compare it with zero-forcing in Fig. 2, where zero-forcing achieves a sum-SDoF lower bound of $2\alpha\min\{[M-N]^+,N\}$ given in \cite{50}. Compared with trivial design (i.e., zero-forcing), our S-RS achieves substantial performance gain. In particular, except perfect CSIT case, our S-RS outperforms the zero-forcing in all antenna configurations. This is because we leverage the power of common message by RS. If CSIT is perfect, our S-RS beats zero-forcing when $M < 2N$, and achieves the same performance as zero-forcing when $2N \le M$. This shows our secure rate splitting is useful even the CSIT is perfect. Moreover, Fig. 2 shows that increasing $M/N$ will not always elevate the achieved sum-SDoF, as the achieved sum-SDoF in \eqref{Result1} saturates since $2 \le M/N$.
 
 \textbf{Corollary 1}: For the two-user MIMO BCCM with imperfect CSIT and a jammer, defined in Section-II, the sum-SDoF achieved by S-RS, i.e., \eqref{Result1}, is SDoF-optimal.
 
  \begin{IEEEproof}
 	Please refer to Appendix A.
 \end{IEEEproof}
 
 \textbf{Corollary 2}: We consider the $K$-user MIMO BCCM with imperfect CSIT and a jammer, where the number of antennas at the transmitter is not less than the sum of antennas at the receiver. The sum-SDoF lower bound achieved by the S-RS is given as follows:
 \begin{equation}
 \text{Sum-SDoF} \ge (1 -\alpha)N + K\alpha N, \qquad M \ge KN.  \label{Result2}
 \end{equation}
 
 \begin{figure}[t]
 	\centering
 	\includegraphics[width=3.3in]{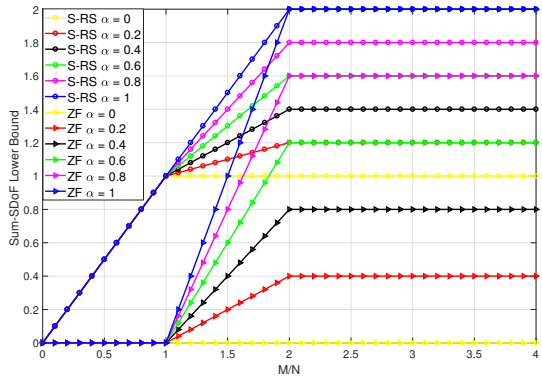}
 	\caption{Comparison of S-RS (S-RS) with zero-forcing (ZF).}
 \end{figure}
 \begin{figure}[t]
 	\centering
 	\includegraphics[width=3.15in]{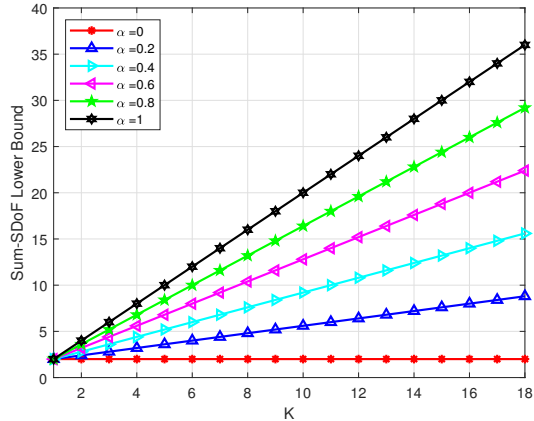}
 	\caption{$K$-user MIMO broadcast channel with imperfect CSIT and a jammer.}
 \end{figure}
 
 \begin{IEEEproof}
 	We extend the settings in \cite{50}, where each receiver can have multiple antennas and the number of transmit antennas is not less than sum of all receive antenna. As such, the proof is similar to that in \cite{50}, and is omitted for simplicity.
 \end{IEEEproof}
 
 In Fig. 3, we illustrate \eqref{Result2}, i.e., Proposition 1, by setting $N=2$. Fig. 3 shows the sum-SDoF lower bound increases with the number of receivers $K$ and CSIT quality $\alpha$. It can be found that the increment by enlarging CSIT quality $\alpha$ becomes greater when  the number of receivers $K$ increase.

\section{The Proposed Secure Rate-Splitting Design}


\textit{Design Principle of S-RS}:
In order to guarantee security and obtain superior transmission efficiency, this design splitting the message into private and common messages with 1-layer rate splitting \cite{50}. 
More specifically, the confidential message for the receiver $i=1,2$ is split into private message $W_{p,i}$ and common message $W_{c,i}$. According to \cite{50}, a common message comprises of packing the common parts of individual messages encoded by a public codebook shared by two receivers, namely $W_c = \{W_{c,1},W_{c,2}\}$.
As such, the transmission of private message is protected by zero-forcing beamforming, while the transmission of common message is protected by jamming from the jammer. The jamming signals are carefully designed to protect the common message transmission, where a jamming codebook in \cite{50} is utilized. This jamming code is designed to mask the undesired common message. 


\subsection{$M \le N$ Antenna Configurations}

As the receiver has more antennas than that of the transmitter, we cannot leverage the zero-forcing beamforming technique\cite{8}. Therefore, the SDoF for private messages is zero.  For common messages,  the design is given as follows: The jammer employs a jamming codebook to generate jamming data stream. The jamming data stream for protecting common message $W_{c,1}$ is denoted by $\textbf{u}_{1} \in \mathbb{C}^M$, which is encoded by the jamming codebook $\{0,V_{c,2}\}$. Meanwhile, the jamming data stream for protecting common message $W_{c,2}$ is denoted by $\textbf{u}_{2} \in \mathbb{C}^M$, which is encoded by the jamming codebook $\{V_{c,1},0\}$. As such, the receiver 1 will have $\{W_{c,1},W_{c,2}+V_{c,2}\}$, and the receiver 2 will have $\{W_{c,1}+V_{c,1},W_{c,2}\}$.  The jamming data steams $\textbf{u}_1$ and $\textbf{u}_2$ transmitted from the jammer are precoded by $\textbf{W}_{c,1} \in \mathbb{C}^{J \times M}$ and $\textbf{W}_{c,2} \in \mathbb{C}^{J \times M}$, respectively. The transmit power allocated to each jamming and common messages is scaled as $\mathcal{O}(P^{M\alpha})$ and $\mathcal{O}(P^M)$, respectively. The transmit signal at the jammer is given by
\begin{equation}
	\textbf{x}_h = \textbf{W}_{c,1}\textbf{u}_1 + \textbf{W}_{c,2}\textbf{u}_2.
\end{equation}
The transmit signal at the transmitter is given by 
\begin{equation}
	\textbf{x}_t = (\textbf{P}_{c,1} + \textbf{P}_{c,2})\textbf{s}_c,
\end{equation}
where the data stream $\textbf{s}_c \in \mathbb{C}^{M}$ is encoded from common message $W_c$, the precoder for common message and receivers 1 and 2 are denoted by $\textbf{P}_{c,1} \in \mathbb{C}^{M \times M}$ and  $\textbf{P}_{c,2} \in \mathbb{C}^{M \times M}$, respectively.  Thus, the received signal at the receiver 1 is expressed as 
\begin{equation}
	\textbf{y}_1 = \textbf{H}_1\textbf{P}_{c,1}\textbf{s}_c + \textbf{H}_1\textbf{P}_{c,2}\textbf{s}_c + \textbf{G}_1\textbf{W}_{c,1}\textbf{u}_1 + \textbf{G}_1\textbf{W}_{c,2}\textbf{u}_2 + \textbf{n}_1, \label{R11}
\end{equation} 
and the received signal at the receiver 2 is expressed as 
\begin{equation}
	\textbf{y}_2 = \textbf{H}_2\textbf{P}_{c,1}\textbf{s}_c + \textbf{H}_2\textbf{P}_{c,2}\textbf{s}_c + \textbf{G}_2\textbf{W}_{c,1}\textbf{u}_1 + \textbf{G}_2\textbf{W}_{c,2}\textbf{u}_2 + \textbf{n}_2, \label{R12}
\end{equation} 
Since the channel is partitioned into the perfect CSI part $\hat{\textbf{H}}_i$ and the unknown CSI part $\tilde{\textbf{H}}_i$, we can further re-write \eqref{R11} and \eqref{R12} into the following:
\begin{subequations}
\begin{eqnarray}
	&& \textbf{y}_1 = \hat{\textbf{H}}_1\textbf{P}_{c,1}\textbf{s}_c + \tilde{\textbf{H}}_1\textbf{P}_{c,1}\textbf{s}_c + \hat{\textbf{H}}_1\textbf{P}_{c,2}\textbf{s}_c + \tilde{\textbf{H}}_1\textbf{P}_{c,2}\textbf{s}_c, \nonumber \\
	&& \qquad + \textbf{G}_1\textbf{W}_{c,1}\textbf{u}_1 + \textbf{G}_1\textbf{W}_{c,2}\textbf{u}_2  + \textbf{n}_1, \\
		&& \textbf{y}_2 = \hat{\textbf{H}}_2\textbf{P}_{c,1}\textbf{s}_c + \tilde{\textbf{H}}_2\textbf{P}_{c,1}\textbf{s}_c + \hat{\textbf{H}}_2\textbf{P}_{c,2}\textbf{s}_c + \tilde{\textbf{H}}_2\textbf{P}_{c,2}\textbf{s}_c, \nonumber \\
	&& \qquad + \textbf{G}_2\textbf{W}_{c,1}\textbf{u}_1 + \textbf{G}_2\textbf{W}_{c,2}\textbf{u}_2  + \textbf{n}_2.
\end{eqnarray}
\end{subequations}
The aim of the S-RS precoder design is to null the information leakage signals to the undesired receiver and align the jamming codebook with the desired receiver. To this end, the precoder can be designed as follows: 
\begin{subequations}
	\begin{eqnarray}
		\text{Nulling:} && \textbf{G}_1\textbf{W}_{c,2} = \textbf{0}, \\	
		&& \textbf{G}_2\textbf{W}_{c,1} = \textbf{0}, \\
		\text{Alignment:} && \hat{\textbf{H}}_1\textbf{P}_{c,1} = \textbf{G}_1\textbf{W}_{c,1}, \\
		&& \hat{\textbf{H}}_2\textbf{P}_{c,2} = \textbf{G}_2\textbf{W}_{c,2}.
	\end{eqnarray}
\end{subequations}
Thereafter, the received signals can be simplified as follows:
\begin{subequations}
	\begin{eqnarray}
&&	 \textbf{y}_1 = (\hat{\textbf{H}}_1\textbf{P}_{c,1} + \tilde{\textbf{H}}_1\textbf{P}_{c,1}  + \tilde{\textbf{H}}_1\textbf{P}_{c,2} + \hat{\textbf{H}}_1\textbf{P}_{c,2})\textbf{s}_c   \nonumber \\
&& \qquad + \hat{\textbf{H}}_1\textbf{P}_{c,1}\textbf{u}_1  + \textbf{n}_1,\\ 
&&	 \textbf{y}_2 =  (\hat{\textbf{H}}_2\textbf{P}_{c,2}+\tilde{\textbf{H}}_2\textbf{P}_{c,1}  + \tilde{\textbf{H}}_2\textbf{P}_{c,2}+\hat{\textbf{H}}_2\textbf{P}_{c,1})\textbf{s}_c \nonumber \\
&& \qquad + \hat{\textbf{H}}_2\textbf{P}_{c,2}\textbf{u}_2 + \textbf{n}_2, 
		\end{eqnarray}
	\end{subequations}
Under the above expression, the SDoF can be analyzed below.

Firstly, in presence of the eavesdropping at receiver 2, the wiretapped SINR for common message $W_{c,1}$ can be given by
\begin{equation}
	\gamma_{2,c} = \frac{\|\textbf{H}_2\textbf{P}_{c,1}  + \textbf{H}_2\textbf{P}_{c,2}\|^2}{\|\hat{\textbf{H}}_2\textbf{P}_{c,2}\|^2 + \|\textbf{n}_2\|^2}.
\end{equation}
Since $\|\hat{\textbf{H}}_2\textbf{P}_{c,2}\|^2$ scales as $\mathcal{O}(P^M)$ and $\|\hat{\textbf{H}}_2\textbf{P}_{c,2}+\tilde{\textbf{H}}_2\textbf{P}_{c,1}  + \tilde{\textbf{H}}_2\textbf{P}_{c,2}+\hat{\textbf{H}}_2\textbf{P}_{c,1}\|^2$ scale as $\mathcal{O}(P^M + P^{M(1 - \alpha)}))$, the information leakage rate $\log(1+\gamma_{2,c})$ is $0$. Likewise, the information leakage rate $\log(1+\gamma_{1,c})$ is $0$ as well. On the other hand, the SNR of information rate for common message $W_c$ can be given by
\begin{equation}
	\eta_{1,c} = \frac{\|\textbf{H}_1\textbf{P}_{c,1} + \tilde{\textbf{H}}_1\textbf{P}_{c,1}\|^2}{ \|\textbf{n}_1\|^2}.
\end{equation}
Since both $\|\textbf{H}_1\textbf{P}_{c,1} + \tilde{\textbf{H}}_1\textbf{P}_{c,1}\|^2$ scales as $\mathcal{O}(P^M + P^{M(1-\alpha)})$, the SDoF of common message is $M$. Therefore, the sum-SDoF lower bound for this case is $M$.

\subsection{$N < M \le 2N$ Antenna Configurations}

As the transmitter has more antennas than that of receiver, we are able to utilize the zero-beamforming to deliver private message. The transmit design for common message is the similar to that in $M \le N$ Case. The transmit design for private messages is given below. The private messages $W_1$ and $W_2$ are encoded into $\textbf{s}_1 \in M-N$ and $\textbf{s}_2 \in M-N$, respectively, which are precoded with $\textbf{P}_1 \in \mathbb{C}^{M \times M-N}$ and $\textbf{P}_2 \in \mathbb{C}^{M \times M-N}$, respectively. The transmit power allocated to each private, jamming, and common messages is scaled as $\mathcal{O}(P^{(M-N)\alpha})$, $\mathcal{O}(P^{N\alpha})$, and $\mathcal{O}(P^N)$, respectively. The transmit signal at the jammer is given by  
\begin{equation}
	\textbf{x}_h = \textbf{W}_{c,1}\textbf{u}_1 + \textbf{W}_{c,2}\textbf{u}_2,
\end{equation}
where jamming data streams $\textbf{u}_1 \in \mathbb{C}^N$ and $\textbf{u}_2 \in \mathbb{C}^N$ are precoded with $\textbf{W}_{c,1} \in \mathbb{C}^{J \times N}$ and $\textbf{W}_{c,2} \in \mathbb{C}^{J \times N}$, respectively. The transmit signal at the transmitter is given by 
\begin{equation}
	\textbf{x}_t = (\textbf{P}_{c,1} + \textbf{P}_{c,2})\textbf{s}_c + \textbf{P}_1\textbf{s}_1 + \textbf{P}_2\textbf{s}_2,
\end{equation}
where common data stream $\textbf{s}_c \in \mathbb{C}^{N}$ are joint precoded with $\textbf{P}_{c,1} \in \mathbb{C}^{M \times N}$ and $\textbf{P}_{c,2} \in \mathbb{C}^{M \times N}$, the private data streams $\textbf{s}_1$ and $\textbf{s}_2$ are precoded with $\textbf{P}_1 $ and $\textbf{P}_2$, respectively. Thus, the received signal at the receiver 1 is expressed as
\begin{eqnarray}
	&& \textbf{y}_1 = \textbf{H}_1\textbf{P}_{c,1}\textbf{s}_c + \textbf{H}_1\textbf{P}_{c,2}\textbf{s}_c + \textbf{H}_1\textbf{P}_{1}\textbf{s}_1 + \textbf{H}_1\textbf{P}_{2}\textbf{s}_2 \nonumber \\
	&& \qquad + \textbf{G}_1\textbf{W}_{c,1}\textbf{u}_1 + \textbf{G}_1\textbf{W}_{c,2}\textbf{u}_2 + \textbf{n}_1, \label{R21}
\end{eqnarray}
and the received signal at the receiver 2 is expressed as 
\begin{eqnarray}
	&& \textbf{y}_2 = \textbf{H}_2\textbf{P}_{c,1}\textbf{s}_c + \textbf{H}_2\textbf{P}_{c,2}\textbf{s}_c + \textbf{H}_2\textbf{P}_{1}\textbf{s}_1 + \textbf{H}_2\textbf{P}_{2}\textbf{s}_2 \nonumber \\
	&& \qquad + \textbf{G}_2\textbf{W}_{c,1}\textbf{u}_1 + \textbf{G}_2\textbf{W}_{c,2}\textbf{u}_2 + \textbf{n}_2, \label{R22}
\end{eqnarray}
Since the channel is partitioned into the perfect CSI part $\hat{\textbf{H}}_i$ and the unknown CSI part $\tilde{\textbf{H}}_i$, we can further re-write \eqref{R21} and \eqref{R22} into the following:
\begin{subequations}
	\begin{eqnarray}
		&& \textbf{y}_1 = \hat{\textbf{H}}_1\textbf{P}_{c,1}\textbf{s}_c + \tilde{\textbf{H}}_1\textbf{P}_{c,1}\textbf{s}_c + \hat{\textbf{H}}_1\textbf{P}_{c,2}\textbf{s}_c + \tilde{\textbf{H}}_1\textbf{P}_{c,2}\textbf{s}_c \nonumber \\
		&& \qquad  + \hat{\textbf{H}}_1\textbf{P}_{1}\textbf{s}_1  + \tilde{\textbf{H}}_1\textbf{P}_{1}\textbf{s}_1  +  \hat{\textbf{H}}_1\textbf{P}_{2}\textbf{s}_2  + \tilde{\textbf{H}}_1\textbf{P}_{2}\textbf{s}_2  \nonumber \\
		&& \qquad +\textbf{G}_1\textbf{W}_{c,1}\textbf{u}_1 + \textbf{G}_1\textbf{W}_{c,2}\textbf{u}_2  + \textbf{n}_1, \\
		&& \textbf{y}_2 = \hat{\textbf{H}}_2\textbf{P}_{c,1}\textbf{s}_c + \tilde{\textbf{H}}_2\textbf{P}_{c,1}\textbf{s}_c + \hat{\textbf{H}}_2\textbf{P}_{c,2}\textbf{s}_c + \tilde{\textbf{H}}_2\textbf{P}_{c,2}\textbf{s}_c \nonumber \\
		&& \qquad  + \hat{\textbf{H}}_2\textbf{P}_{1}\textbf{s}_1  + \tilde{\textbf{H}}_2\textbf{P}_{1}\textbf{s}_1  +  \hat{\textbf{H}}_2\textbf{P}_{2}\textbf{s}_2  + \tilde{\textbf{H}}_2\textbf{P}_{2}\textbf{s}_2  \nonumber \\
		&& \qquad +\textbf{G}_2\textbf{W}_{c,1}\textbf{u}_1 + \textbf{G}_2\textbf{W}_{c,2}\textbf{u}_2  + \textbf{n}_2.
	\end{eqnarray}
\end{subequations}
The aim of the S-RS precoder design is to null the information leakage signals to the undesired receiver and align the jamming codebook with the desired receiver. To this end, the precoder can be designed as follows: 
\begin{subequations}
	\begin{eqnarray}
		\text{Nulling:} && \textbf{G}_1\textbf{W}_{c,2} = \textbf{0}, \\	
		&& \textbf{G}_2\textbf{W}_{c,1} = \textbf{0}, \\		
		&& \hat{\textbf{H}}_1\textbf{P}_{2} = \textbf{0}, \\
		&& \hat{\textbf{H}}_2\textbf{P}_{1} = \textbf{0}, \\
		\text{Alignment:} && \hat{\textbf{H}}_1\textbf{P}_{c,1} = \textbf{G}_1\textbf{W}_{c,1}, \\
		&& \hat{\textbf{H}}_2\textbf{P}_{c,2} = \textbf{G}_2\textbf{W}_{c,2}.
	\end{eqnarray}
\end{subequations} Thus, due to the rank of both $\hat{\textbf{H}}_1$ and $\hat{\textbf{H}}_2$ is $\min\{M,N\}$, the rank of both $\textbf{P}_{c,1}$ and $\textbf{P}_{c,2}$ is $M-N$. Thereafter, the received signals can be simplified as follows:
\begin{subequations}
	\begin{eqnarray}
		&&	 \textbf{y}_1 = (\hat{\textbf{H}}_1\textbf{P}_{c,1} + \tilde{\textbf{H}}_1\textbf{P}_{c,1}  + \tilde{\textbf{H}}_1\textbf{P}_{c,2} + \hat{\textbf{H}}_1\textbf{P}_{c,2})\textbf{s}_c   \nonumber \\
		&& \qquad + \hat{\textbf{H}}_1\textbf{P}_{c,1}\textbf{u}_1  + (\hat{\textbf{H}}_1\textbf{P}_{1} + \tilde{\textbf{H}}_1\textbf{P}_{1})\textbf{s}_1 \nonumber \\
		&& \qquad  + \tilde{\textbf{H}}_1\textbf{P}_{2}\textbf{s}_2 + \textbf{n}_1,\\ 
		&&	 \textbf{y}_2 =  (\hat{\textbf{H}}_2\textbf{P}_{c,2}+\tilde{\textbf{H}}_2\textbf{P}_{c,1}  + \tilde{\textbf{H}}_2\textbf{P}_{c,2}+\hat{\textbf{H}}_2\textbf{P}_{c,1})\textbf{s}_c \nonumber \\
		&& \qquad + \hat{\textbf{H}}_2\textbf{P}_{c,2}\textbf{u}_2 + (\hat{\textbf{H}}_2\textbf{P}_{2} + \tilde{\textbf{H}}_2\textbf{P}_{2})\textbf{s}_2 \nonumber \\
		&& \qquad + \tilde{\textbf{H}}_2\textbf{P}_{1}\textbf{s}_1  + \textbf{n}_2, 
	\end{eqnarray}
\end{subequations}
Henceforth, the common data steam is firstly decoded by treating interference as noise. After decoding of $\textbf{s}_c$, its impact is canceled. Then, the private data steams $\textbf{s}_1$ and $\textbf{s}_2$ are decoded at receivers 1 and 2, respectively. Under the above decoding procedure, the SDoF can be analyzed below.

Firstly, in presence of the eavesdropping at receiver 2, the wiretapped SINR for common message $W_{c,1}$ can be given by
\begin{equation}
	\gamma_{2,c} =	\frac{\|\textbf{H}_2\textbf{P}_{c,1} + \textbf{H}_2\textbf{P}_{c,2}\|^2}{\|\hat{\textbf{H}}_2\textbf{P}_{c,2}\|^2 + \|\hat{\textbf{H}}_2\textbf{P}_{2} + \tilde{\textbf{H}}_2\textbf{P}_{2}\|^2 + \|\tilde{\textbf{H}}_2\textbf{P}_{1}\|^2 + \|\textbf{n}_2\|^2}.
	\end{equation}
Since $\|\hat{\textbf{H}}_2\textbf{P}_{c,2}\|^2$ scales as $\mathcal{O}(P^N)$, $\|\textbf{H}_2\textbf{P}_{2}\|^2$ scales as $\mathcal{O}(P^{(M-N)\alpha})$, and $\|\textbf{H}_2\textbf{P}_{c,2}+ \textbf{H}_2\textbf{P}_{c,1}\|^2$ scale as $\mathcal{O}(P^N + P^{N(1 - \alpha)})$, the information leakage rate $\log(1+\gamma_{2,c})$ is $0$. Likewise, the information leakage rate $\log(1+\gamma_{1,c})$ is $0$ as well. After cancellation of the impact of common message $W_c$, the wiretapped SINR for private message $W_{p,1}$, in presence of the eavesdropping at receiver 2, can be given by
\begin{equation}
	\gamma_{2,p} = \frac{\|\tilde{\textbf{H}}_2\textbf{P}_{1}\|^2}{ \|\textbf{n}_2\|^2},
\end{equation}
Since $\|\tilde{\textbf{H}}_2\textbf{P}_{1}\|^2$ scales as $\mathcal{O}(1)$, the information leakage rate $\log(1 + \gamma_{2,p})$ is 0. Likewise, the the information leakage rate $\log(1 + \gamma_{1,p})$ is 0 as well. On the other hand, the SINR of information rate for common message $W_{c,1}$ can be given by
\begin{equation}
	\eta_{1,c} = \frac{\|\textbf{H}_1\textbf{P}_{c,1}+\tilde{\textbf{H}}_1\textbf{P}_{c,2}\|^2}{\|\textbf{H}_1\textbf{P}_{1}\|^2 + \|\tilde{\textbf{H}}_1\textbf{P}_{2}\|^2 + \|\textbf{n}_1\|^2}.
\end{equation}
Since $\|\textbf{H}_1\textbf{P}_{c,1}  + \tilde{\textbf{H}}_1\textbf{P}_{c,2}\|^2$ scales as $\mathcal{O}(P^N + P^{N(1-\alpha)})$, $\|\textbf{H}_1\textbf{P}_{1}\|^2$ scales as $\mathcal{O}(P^{(M-N)\alpha})$, $\|\tilde{\textbf{H}}_1\textbf{P}_{2}\|^2$ scales as $\mathcal{O}(1)$, the SDoF of common message is $N-(M-N)\alpha$. After cancellation of the impact of common message $W_c$, the SINR of information rate for private message $W_{p,1}$ is given by
\begin{equation}
	\eta_{1,p} = \frac{\|\textbf{H}_1\textbf{P}_{1}\|^2}{ \|\tilde{\textbf{H}}_1\textbf{P}_{2}\|^2 + \|\textbf{n}_1\|^2}.
\end{equation}
Since $\|\textbf{H}_1\textbf{P}_{1}\|^2$ scales as $\mathcal{O}(P^{(M-N)\alpha})$ and $\|\tilde{\textbf{H}}_1\textbf{P}_{2}\|^2$ scales as 
$\mathcal{O}(1)$, the SDoF of private message $W_{p,1}$ is $(M-N)\alpha$. Likewise, the SDoF of private message $W_{p,1}$ is $(M-N)\alpha$ as well. Therefore, the sum-SDoF lower bound for this case is $N-(M-N)\alpha+2\alpha(M-N)$.

\subsection{$2N < M$ Antenna Configurations}

Here, we set $M=2N$. Thus, it degenerates to the one in $N < M \le 2N$ Case. The sum-SDoF lower bound for this case is $N(1 + \alpha)$ by setting $M=2N$ in $N(1-\alpha)+2\alpha(M-N)$.



\section{Conclusion}

We studied S-RS for the two-user MIMO broadcast channel with imperfect CSIT and a multiple-antenna jammer having perfect CSI. Specifically, we designed the transmit precoders to enable the S-RS by joint nulling and aligning the leakage information, regarding to different antenna configurations. Moreover, we revealed that the sum-SDoF achieved by S-RS is optimal and outperformed that by conventional zero-forcing,  which validated the superiority of S-RS in the two-user MIMO broadcast channel.  



\begin{appendices}
	
\section{Proof of Corollary 1}

According to \cite{203,260}, the DoF region of the two-user MIMO broadcast channel with imperfect CSIT is given in (28a)-(28c)
\begin{figure*}
	\begin{subequations}
			\begin{eqnarray}
			&&	\text{DoF of Receiver 1} \le \min\{M,N\}, \\
			&&	\text{DoF of Receiver 2} \le \min\{M,N\}, \\
			&&	\text{DoF of Receiver 1} + \text{DoF of Receiver 2} \le \min\{M,N\} + \alpha(\min\{M,2N\}-\min\{M,N\}). 
		\end{eqnarray}
	\end{subequations}
 \hrule
\end{figure*}
The sum-DoF can be derived from (28a)-(28c), i.e., 
\begin{eqnarray}
\text{Sum-DoF} = \begin{cases}
 M, & \frac{M}{N} \le 1, \\
N + \alpha(M-N), & 1 < \frac{M}{N} \le 2, \\
N(1+\alpha), & 2 < \frac{M}{N}.
\end{cases}, \label{UBound}
\end{eqnarray}
which is an upper bound of the sum-SDoF of the two-user MIMO broadcast channel with imperfect CSIT and a jammer. Therefore, it can be seen from \eqref{UBound} that the sum-SDoF in Theorem 1 is SDoF-optimal.

\end{appendices}

\bibliographystyle{IEEEtran}
\bibliography{ICCWCRef}
 
\end{document}